\documentclass{aa}
\usepackage{epsfig}
\usepackage{natbib}
\usepackage{graphicx}
\begin{document}
\newcommand{\av}[1]{\ensuremath{\langle#1\rangle}}
\newcommand{\dd}{\ensuremath{\mathrm{d}}}
\newcommand{\bea}{\begin{eqnarray}}    
\newcommand{\eea}{\end{eqnarray}}      
\newcommand{\be}{\begin{equation}}
\newcommand{\ee}{\end{equation}}
\newcommand{\bef}{\begin{figue}}
\newcommand{\eef}{\end{figure}}
\newcommand{\etal}{et al.}
\newcommand{\kms}{\,{\rm km}\;{\rm s}^{-1}}
\newcommand{\hubunits}{\,\kms\;{\rm Mpc}^{-1}}
\newcommand{\hmpc}{\,h^{-1}\;{\rm Mpc}}
\newcommand{\hkpc}{\,h^{-1}\;{\rm kpc}}
\newcommand{\msun}{M_\odot}
\newcommand{\K}{\,{\rm K}}
\newcommand{\cm}{{\rm cm}}
\newcommand{\cd}{{\langle n(r) \rangle_p}}
\newcommand{\Mpc}{{\rm Mpc}}
\newcommand{\kpc}{{\rm kpc}}
\newcommand{\xir}{{\xi(r)}}
\newcommand{\xrp}{{\xi(r_p,\pi)}}
\newcommand{\xsirpi}{{\xi(r_p,\pi)}}
\newcommand{\wrp}{{w_p(r_p)}}
\newcommand{\gr}{{g-r}}
\newcommand{\Navg}{N_{\rm avg}}
\newcommand{\Mmin}{M_{\rm min}}
\newcommand{\fiso}{f_{\rm iso}}
\newcommand{\Mr}{M_r}
\newcommand{\rp}{r_p}
\newcommand{\zmax}{z_{\rm max}}
\newcommand{\zmin}{z_{\rm min}}

\def\eg{{e.g.}}
\def\ie{{i.e.}}
\def\spose#1{\hbox to 0pt{#1\hss}}
\def\ltapprox{\mathrel{\spose{\lower 3pt\hbox{$\mathchar"218$}}
\raise 2.0pt\hbox{$\mathchar"10C$}}}
\def\gtapprox{\mathrel{\spose{\lower 3pt\hbox{$\mathchar"218$}}
\raise 2.0pt\hbox{$\mathchar"10E$}}}
\def\inapprox{\mathrel{\spose{\lower 3pt\hbox{$\mathchar"218$}}
\raise 2.0pt\hbox{$\mathchar"232$}}}

\title{Breaking of self-averaging properties and selection effects in
  the Luminous Red Galaxies sample}

\subtitle{}

\author{Francesco Sylos Labini \inst{1,2}}

\titlerunning{Self-averaging properties and selection in the LRG sample}
\authorrunning{Sylos Labini}

\institute{ 
Centro Studi e Ricerche Enrico Fermi, Via Panisperna 89 A, 
Compendio del Viminale, 00184 Rome, Italy
\and Istituto dei Sistemi Complessi CNR, 
Via dei Taurini 19, 00185 Rome, Italy.
}

\date{Received / Accepted}

\abstract{We study the statistical properties of the Luminous Red
  Galaxies sample from the Sloan Digital Sky Survey. In particular we
  test, by determining the probability density function (PDF) of
  galaxy (conditional) counts in spheres, whether statistical
  properties are self-averaging within the sample. We find that there
  are systematic differences in the shape of the PDF and in the
  location of its peak, signaling that there are major systematic
  effects in the data which make the estimation of volume average
  quantities unreliable within this sample. We discuss that these
  systematic effects are related to the fluctuating behavior of the
  redshift counts which can be originated by intrinsic fluctuations in
  the galaxy density field or by observational selection effects. The
  latter possibility implies that more than $20 \%$ of the galaxies
  have not been observed and that such a selection should not be a
  smooth function of redshift.
 } 
  
\maketitle

\section{Introduction} 

The sample of Luminous Red Galaxies (LRG) \citep{lrg} from the Sloan
Digital Sky Survey (SDSS) \citep{sdss} is considered to be the best
sample to study galaxy large scale structures. One of the main
features observed in the LRG sample from the final data release of the
SDSS \citep{dr7}, is that the number density as a function of redshift
$n(z)$, usually called the ``selection function'', presents a complex
behavior.  Given that, by construction, the LRG sample should be
volume-limited \citep{lrg,zehavi,kazin} the behavior of $n(z)$ is
expected to be constant if galaxy distribution is close to uniform (up
to Poisson noise and radial clustering).  It is instead observed that
the LRG sample $n(z)$ shows an irregular and not constant behavior.
An explanation that it is usually given for this result
\citep{zehavi,kazin}, is that the LRG sample is ``quasi'' volume
limited, in that it does not show a constant $n(z)$. Thus, the
features in $n(z)$ are absorbed in the properties of a selection
function, which is unknown {\it a priori}, but that it is defined {\it
  a posteriori} as the difference between an almost constant $n(z)$
and the behavior observed.  Clearly this explanation is unsatisfactory
as it is given {\it a posteriori} and no independent tests have been
provided to corroborate the hypothesis that an important observational
selection effect occurs in the data, other than the behavior of $n(z)$
itself. Indeed a different possibility is that the behavior of $n(z)$
is determined by intrinsic fluctuations in the distribution of
galaxies and not by selection effects\footnote{Note that for $z>0.36$
  there is instead a clear selection effect due to the ``passage of
  the 4000 A break into the $r$ band'' \citep{kazin}.  This is shown
  by a smooth redshift-dependent decrease of the redshift counts. For
  this reason we limit our analysis to $z = 0.36$. }.

We note that if (unknown) selection effects are considered to play a
major role in this sample, the question that should be addressed
concerns the quality of the data. In the original paper by \citet{lrg}
it is stated that the sample is volume-limited up to $z\sim 0.38$,
modulo minor effects due to K+$e$ corrections (i.e. K-corrections and
luminosity evolution).  To support that this is the case, $n(z)$ was
found, in the early SDSS LRG data (see their Figs.12-13), to be close
to flat. However \citet{eisenstein}, by considering a much larger LRG
sample, noted that the comoving number density displays a close to a
constant behavior, ``although fluctuations are reaching about the
$30\%$ peak-to-peak'' (see their Fig.1). More recently, by considering
the LRG sample with the same selection in magnitude and redshift as
the one considered by \citet{eisenstein}, but from the final SDSS data
release which is larger by a factor $\sim 3$ in volume, \citet{kazin}
(see their Fig.1) noted that there are non-negligible variations in
$n(z)$, that, as mentioned above, make difficult to consider this
sample as a purely volume limited one.

In this paper we determine the probability density function (PDF) of
conditional galaxy counts in spheres \footnote{Conditional properties
  are local, and they do not require the measurement of global
  statistical quantities on the sample scale. They are well defined
  both for spatially uniform and inhomogeneous distributions, while
  unconditional statistical quantities are well defined only for
  spatially homogeneous systems \citep{book}.}.  In practice we study
the statistical properties of the variable $N_i(r)$, which is defined
to be the number of galaxies in a sphere a radius $r$ centered on the
$i^{th}$ galaxy.  This has the advantage to provide a characterization
of conditional fluctuations which is not affected by the geometrical
constraints of the sample as $n(z)$, which instead explores different
volumes at different redshifts. In addition, by measuring the PDF in
different regions of the sample we can make a clear test to check
whether fluctuations are self-averaging inside the sample. This is the
fundamental property that fluctuations are required to have when an
average quantity is measured in a given sample.

Indeed a spatial average is meaningful, in the sense that it provides
an estimation of the ensemble average property of the given
statistics, only when fluctuations are stationary inside the sample
\citep{sdss_aea,copernican}.  The breaking of self-averaging
properties can be generally due to two different reasons. On the one
hand when a distribution is not statistically translational and/or
rotational invariant, then the fluctuation properties depend on the
specific position of the volume in which they are measured. For
instance, if the distribution is spherically symmetric around a point,
then the relevant role is played by the distance $R$ from the center:
fluctuations at different distances $R$ have different properties and
for this reason any volume average quantity, computed in regions large
enough so that statistical properties (i.e., the local density) change
significantly, do not give an useful information about the real
properties of the distribution. This same situation occurs when a
distance-dependent observational selection effect occurs in the data
as it can mimic the break of translational invariance.

On the other hand self-averaging properties can be broken when
fluctuations, in a given sample, are too extended in space and have
too large amplitude \citep{sdss_aea,sdss_epl}.  For instance when
there are a few large scale structures in given volume, or even a
single one, which dominates the distribution then it is not possible
to get a meaningful estimation of an average quantity at large enough
scales in the sample. This occurs precisely because the sample volume
is not large enough to average between different large enough
structures. This is a systematic effect that sometimes is refereed to
as cosmic variance \citep{saslaw2010} but that is more appropriately
defined as breaking of self-averaging properties \citep{sdss_aea}, as
the concept of variance (which involves already the computation of an
average quantity) maybe without statistical meaning in the
circumstances described above.

The breaking of self-averaging properties was found to occur in
several volume-limited samples of the main galaxy sample (MGS) of the
DR6-SDSS \citep{sdss_aea}.  Indeed, it was found that at large enough
scales, i.e. $r>30$ Mpc/h, the PDF of conditional fluctuations shows
systematic differences when it is measured in volumes located in
different positions in the sample. That this corresponds to the
breaking of self-averaging properties and not the the breaking of
statistical isotropy and/or homogeneity (for intrinsic or
observational reasons) is shown by two facts. The first is that the
difference between the PDF, measured into two volumes located at small
and high redshifts, is not systematically the same. That is, sometimes
the PDF measured at low redshift is shifted toward smaller values and
sometimes toward higher values. As discussed by \citet{copernican} the
breaking of statistical translational invariance, as well as the
effect of a redshift dependent selection effect, would be signed by a
difference in the PDF at low and high redshift always in the same
direction. Secondly, when the larger volumes provided by DR7-SDSS was
considered \citep{tibor}, the PDF did not show the same difference:
this is a clear indication that the breaking of self-averaging
properties is due to a systematic volume-dependent effect. In
addition, when self-averaging properties were found to be satisfied
the PDF was found to be nicely fitted by a Gumbel function and its
first and second moments, the average conditional density and the
conditional variance, were found to have a scaling behavior as a
function of distance. All these behaviors mark a clear departure from
spatial uniformity \citep{tibor}.

In this paper we perform the test for self-averaging in the LRG
sample.  In Sect.\ref{sstest} we recall the basic statistical elements
which we use in the data analysis.  The main features of the data are
presented in Sect.\ref{data} while in Sect.\ref{result} we show our
results. We discuss our results in Sect.\ref{discussion} by comparing
the results obtained here with the ones measured in the SDSS-MGS by
\citet{loveday,sdss_aea,sdss_epl,saslaw2010} and in the 2 degree field
redshift survey (2dFGRS) by \citet{2df_epl,2df_aea}.  In addition we
critically consider the measurements of the conditional average
density by \citet{hogg}, the assumptions entering in the
determination of the two-point correlation function \citep{kazin} and
the problem underlying the determination of the so-called baryon
acoustic peak \citep{eisenstein,kazin,martinez,sdss_bao}.  Finally we
draw our conclusions in Sect.\ref{conclusions}.

\section{Statistical background}
\label{sstest}

As it was discussed by \citet{sdss_aea,sdss_epl,copernican} a simple
test to determine whether a point distribution is self-averaging in a
given sample of linear size $R_s$ consists in studying the PDF of
conditional galaxy counts in spheres ${\cal N}$ (which contains, in
principle, information about moments of any order) in sub-samples of
linear size $r < R_s$ placed in different and non-overlapping spatial
regions of the sample (that we call $S_1,S_2,...S_N$).  That the
self-averaging property holds is shown by the fact that $P_{S_i}({
  \cal N};r)$ is the same, modulo statistical fluctuations, in the
different sub-samples, i.e.,
\be
\label{selfavtest}
 P_{S_i}({ \cal N};r) \approx P_{S_j}({ \cal N};r) \; \forall i \ne j \; .
\ee

On the other hand, if determinations of $P_{S_i}({ \cal N},r)$ in
different sample regions $S_i$ show {\it systematic} differences then,
as discussed above, the distribution is not self-averaging because of
the presence of non-averaged large scale structures or because it not
statistically translational and/or rotational invariant (this includes
the case in which a redshift dependent observational selection effect
is present).  When Eq.\ref{selfavtest} is {\it not} found to be
satisfied in a given sample then the determinations of the spatial
averages are sample-dependent implying that those statistical
quantities do not represent the asymptotic properties of the given
distribution \citep{sdss_aea,copernican}.

To test self-averaging is necessary to employ statistical quantities
that do not require the assumption of spatial homogeneity inside the
sample and thus avoid the normalization of fluctuations to the
estimation of the sample average \citep{sdss_aea}.  We therefore
consider the PDF of the stochastic variable defined by number of
points $N_i(r)$ contained in a sphere of radius $r$ centered on the
$i^{th}$ point.  This depends on the scale $r$ and on the spatial
position of the $i^{th}$ sphere's center, namely its radial distance
$R_i$ from a given origin (in this case the Earth) and its angular
coordinates $\vec{\alpha}_i$.  Integrating over $\vec{\alpha}_i$ for
fixed radial distance $R_i$, we obtain that $N_i(r)=N(r; R_i)$
\citep{sdss_aea,tibor,copernican}.

In \citet{tibor} we showed that the Gumbel distribution was a good fit
to the data of the SDSS-MGS, marking a clear departure from the
Gaussian behavior expected for spatially uniform systems.  The Gumbel
distribution's PDF, at fixed $r$, is given by (see \citet{tibor} and
references therein)
\begin{equation}
\label{gumbel}
 P(N)= \frac{1}{\beta} 
\exp\left[ - \frac{N-\alpha}{\beta} - 
\exp\left( - \frac{N-\alpha}{\beta} \right) \right] \;.
\end{equation}
The mean and the standard deviation (variance) of the Gumbel
distribution (Eq.\ref{gumbel}), depend on the scale $r$, and are given
by
\begin{equation}
\label{cumu}
 \mu = \alpha + \gamma \beta, \quad 
 \sigma^2 = (\beta\pi)^2/ 6 
\end{equation}
where $\gamma=0.5772\dots$ is the Euler constant.


\section{The data} 
\label{data}

The sample we consider in this paper is one of the samples prepared by
\citet{kazin}.  In particular we focus on the so-called DR7-Dim, while
we do not present results for the other samples, as the one denoted as
bright that contains too few galaxies for a reliable statistical
analysis.

The metric distance $ R(z; \Omega_m, \Omega_\Lambda) $ from the
redshift has been computed by using the cosmological parameters with
values $\Omega_m=0.25$ and $\Omega_\Lambda=0.75$.

Given that we exclude redshifts $z>0.36$ and $z<0.16$, the distance
limits are: $R_{min} = 465 $ Mpc/h and $R_{max} = 1002$ Mpc/h. The
limits in R.A $\alpha$ and Dec. $\delta$ considered are chosen in such
a way that (i) the angular region does not overlap with the irregular
edges of the survey mask and (ii) the sample covers a contiguous sky
area.  Thus we have chosen: $\alpha_{min} = 130^\circ$ and
$\alpha_{max} = 240^\circ$; $\delta_{min} = 0^\circ$ and $\delta_{max}
= 50^\circ$ . The absolute magnitude is constrained in the range $M
\in [-23.2,-21.2]$. With these limits we find $N=41833$ galaxies
covering a solid angle $\Omega=1.471$ sr.

\section{Results}
\label{result}

As discussed in Sect.\ref{sstest} we study the statistical properties
of the random variable $N_i(r)$ measuring the integrated number of
points in a sphere of radius $r$ around $i=1...M(r)$ points. The
number of points over which it is possible to compute this quantity
at the scale $r$, denoted as $M(r)$, depends on the geometrical
constraints of the sample as we consider only spheres that are fully
enclosed in the sample boundaries (see for details 
\citet{sdss_aea}).  At fixed $r$ we make an histogram
of the $M(r)$ values of $N_i(r)$, which thus represents an estimation of the
PDF $P(N;r)$ of conditional fluctuations. 

In Fig.\ref{fig:radialden} we show the behavior of the number density
as a function of distance. There are two main features: (i) a negative
slope between $400$ Mpc/h $<r<$ 800 Mpc/h (i.e., $0.16 < z < 0.28$)
and (ii) a positive slope up to a local peak at $r\sim 950$ Mpc/h
(i.e., $z \sim 0.34$). Additional features are present as peaks in the
$n(r)$ behavior at $r \sim 500$ ($z \sim 0.18$) and $r\sim 650$ Mpc/h
($z \sim 0.23$). Note that if $n(z)$ were constant we would expect a
behavior similar to the one shown by the mock sample extracted from
the Horizon simulation with the same geometry of the real LRG sample
\citep{horizon} (see Fig.\ref{fig:radialden}).  This implies that, by
addressing the behavior of $n(z)$ to unknown selection effects, it is
implicitly assumed that the survey has lost more than the $20 \%$ (for
$n_0= 1.2 \times 10^{-4}$ galaxies for (Mpc/h)$^{-3}$ ) of the
galaxies for observational problems.  This looks improbable
\citep{lrg} although a more careful investigation of the problem must
be addressed. Note also that the deficit of galaxies would not be
explained by a smooth redshift-dependent effect (as it occurs for
$z>0.36$ --- see Fig.1 of \citet{kazin}) rather the selection must be
strongly redshift dependent as the behavior of $n(z)$ is not
monotonic.  These facts point, but do not proof, toward an origin of
the $n(z)$ behavior due to the intrinsic fluctuations in the galaxy
distribution.
\begin{figure}
\begin{center}
\includegraphics*[angle=0, width=0.5\textwidth]{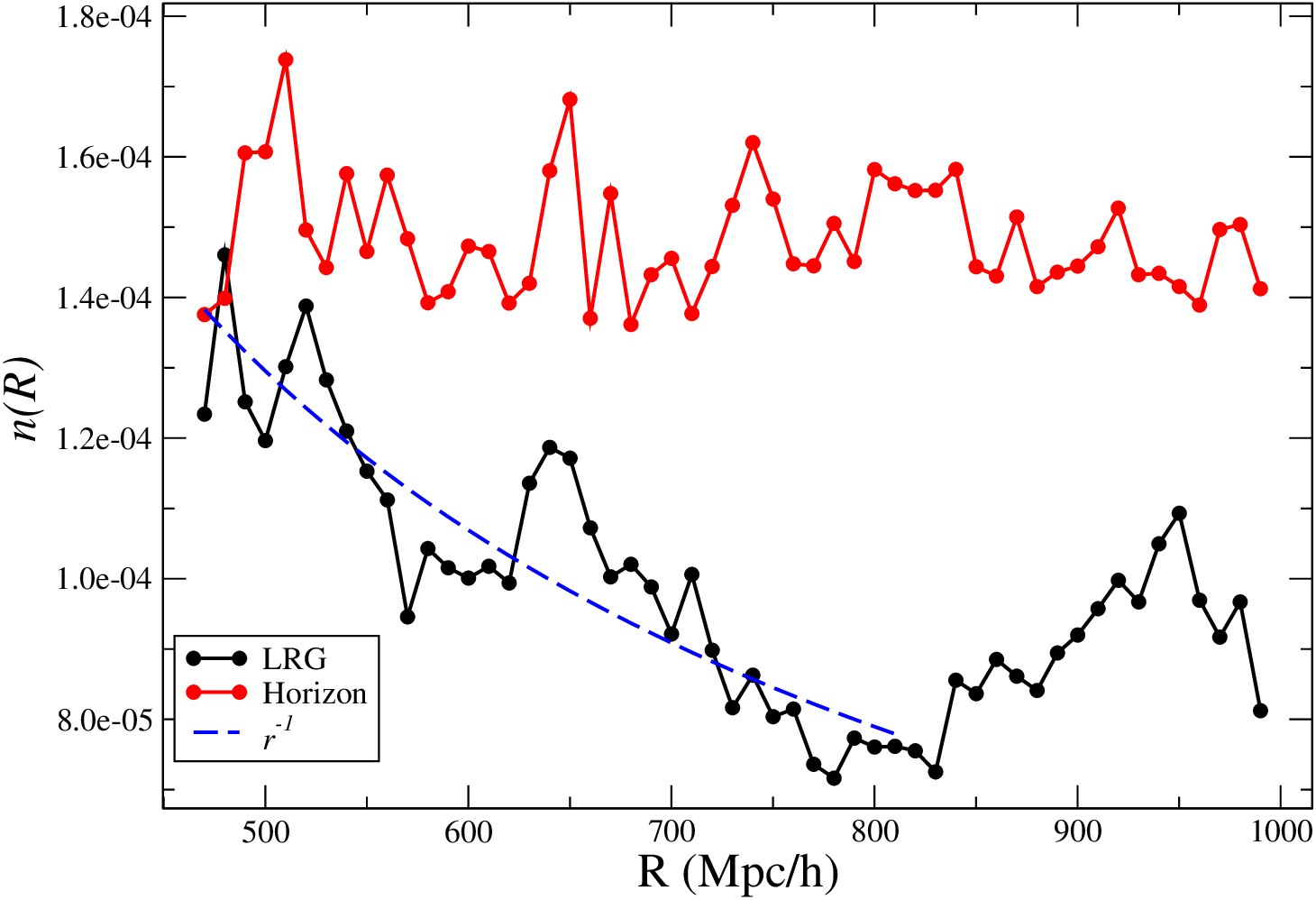}
\end{center}
\caption{Number density as a function of distance for the LRG sample
  and for a mock sample extracted from the Horizon simulations (units
  are in (Mpc/h)$^{-3}$). The blue dashed line decays as $r^{-1}$ and
it is plotted as reference.}
\label{fig:radialden}
\end{figure}

The analysis of fluctuations by studying $n(z)$ shows clearly that the
observed behavior is incompatible with model predictions. However the
redshift distribution provides only a rough analysis of fluctuations,
especially because it is not an average quantity and because it
samples different scales differently as the volume in the different
redshift bins is not the same. For this reason we determine the
statistical properties of the stochastic variable $N_i(r)$ previously
defined, which allows us to compute the full PDF in sub-samples of
equal volume.  In Fig.\ref{fig:Histo1a} we report the PDF at different
scales in the LRG sample, together with the best fit with
Eq.\ref{gumbel}, which provides a reasonably good fit. A
quantification of the difference of the measured PDF from the Gumbel
or Gaussian behaviors is difficult for the problems discussed below.
\begin{figure}
\begin{center}
\includegraphics*[angle=0, width=0.5\textwidth]{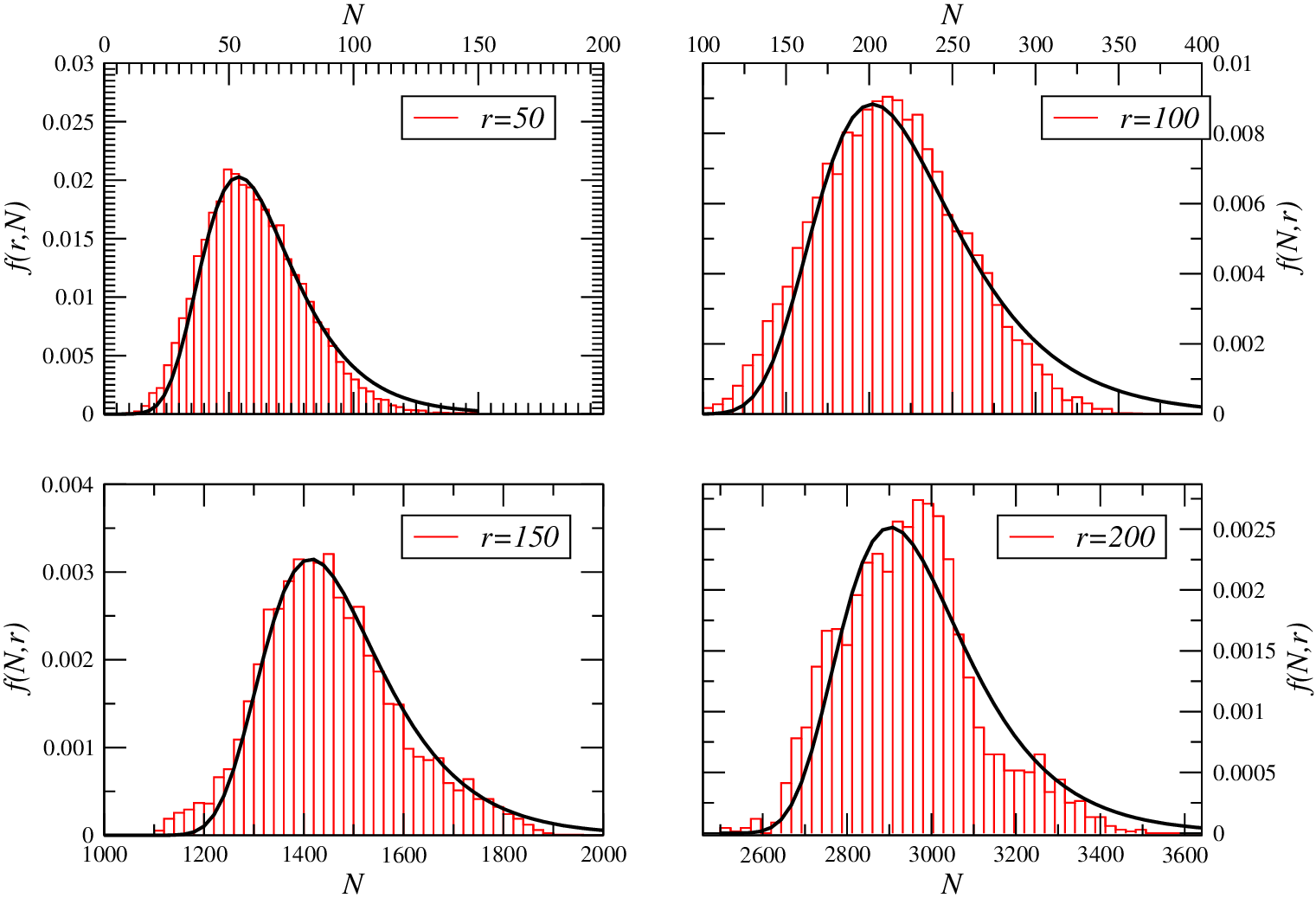}
\end{center}
\caption{PDF for $r=50,100,150,200$ Mpc/h together with the best with
  Eq.\ref{gumbel}. The number of points contributing to the histogram
  is respecitively $M(r)= 26376, 18916, 6645,2829$. }
\label{fig:Histo1a}
\end{figure}

Let us now pass to the self-averaging test described in
Sect.\ref{sstest}.  To this aim we divide the sample into two
non-overlapping regions of same volume, one at low (L) redshifts and
the other at high (H) redshifts. We then measure the PDF $P_L(N;r)$
and $P_H(N;r)$ in the two volumes.  Given that the total number of
points is not very large (i.e., $M(r) \sim 10^4$), in order to improve
the statistics especially for large sphere radii, we allow a partial
overlapping between the two sub-samples, so that galaxies in the L (H)
sub-sample counts also galaxies in the H (L) sub-sample.  This
overlapping clearly can only smooth out differences between $P_L(N;r)$
and $P_H(N;r)$.  Results are shown in
Figs.\ref{fig:pn2r50}-\ref{fig:pn2r200} for $r=50, 100, 150, 200$
Mpc/h respectively. One may note that for $r=50$ Mpc/h the two
determinations are much closer than for lager sphere radii for which
there is actually a noticeable difference in the whole shape of the
PDF \footnote{The Kolmogorov-Smirnoff test \citep{nr} allows, in
  general, to make a statistical test to measure the distance between
  two cumulative PDF. In all cases discussed here, we find that both
  the Gaussian and the Gumbel functions do not fit the data at a
  reasonable level of confidence. In addition when applying the same
  test to study whether the data from the H and S samples are drawn
  from the same distribution, we also find that the answer is in the
  negative.}. The fact that $P_H(N;r)$ is shifted toward smaller
values than $P_L(N;r)$ is related to the decaying behavior of the
redshift counts (see Fig.\ref{fig:radialden}): most of the galaxies at
low redshifts see a relatively larger local density than the galaxies
at higher redshift.

\begin{figure}
\begin{center}
\includegraphics*[angle=0, width=0.5\textwidth]{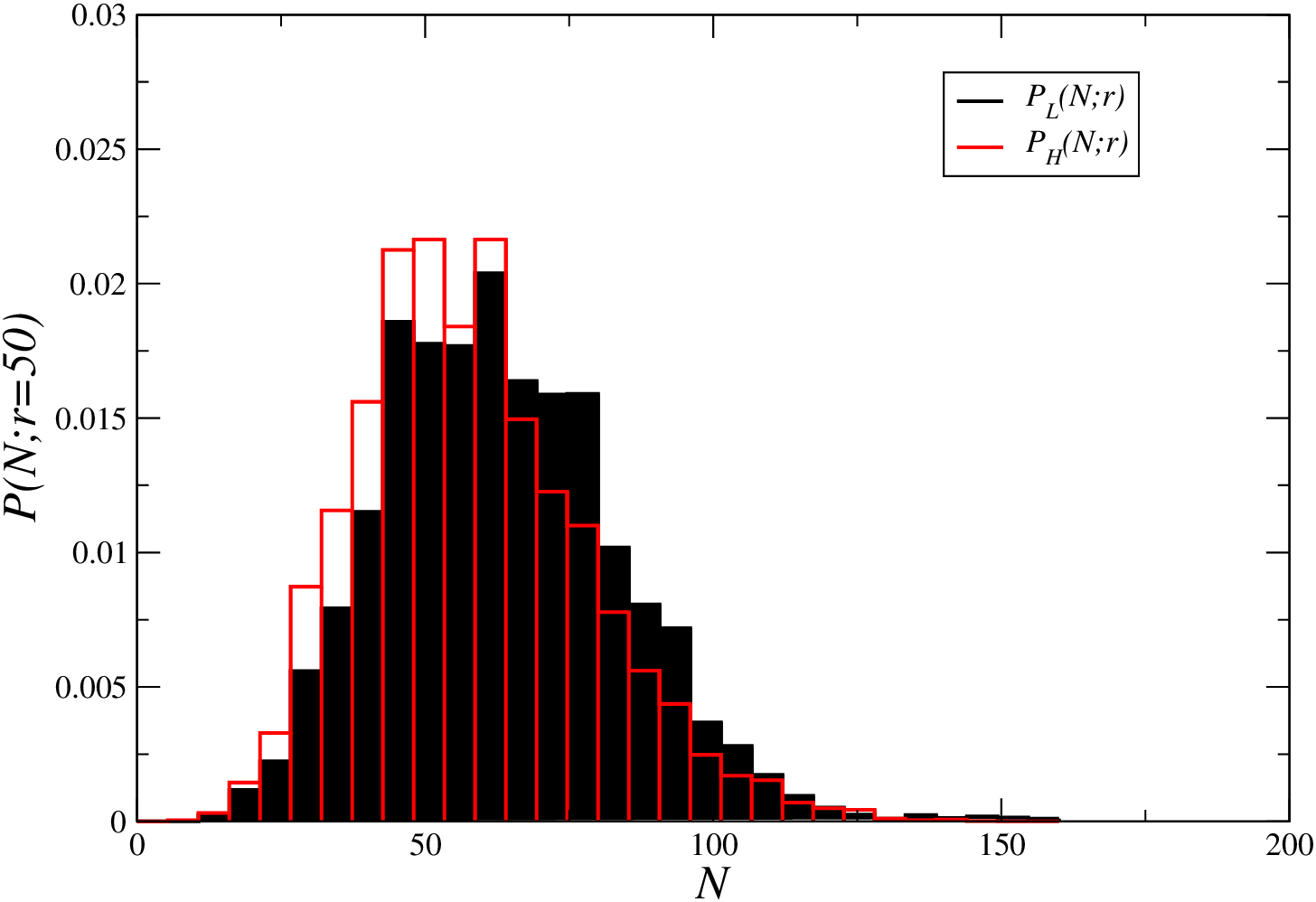}
\end{center}
\caption{PDF for $r=50$ Mpc/h in the low (L) redshifts and the high
  (H) redshifts sample. The number of points contributing to the
  histogram is respecitively for L and H $M(r)= 13277, 13099$.}
\label{fig:pn2r50}
\end{figure}

\begin{figure}
\begin{center}
\includegraphics*[angle=0, width=0.5\textwidth]{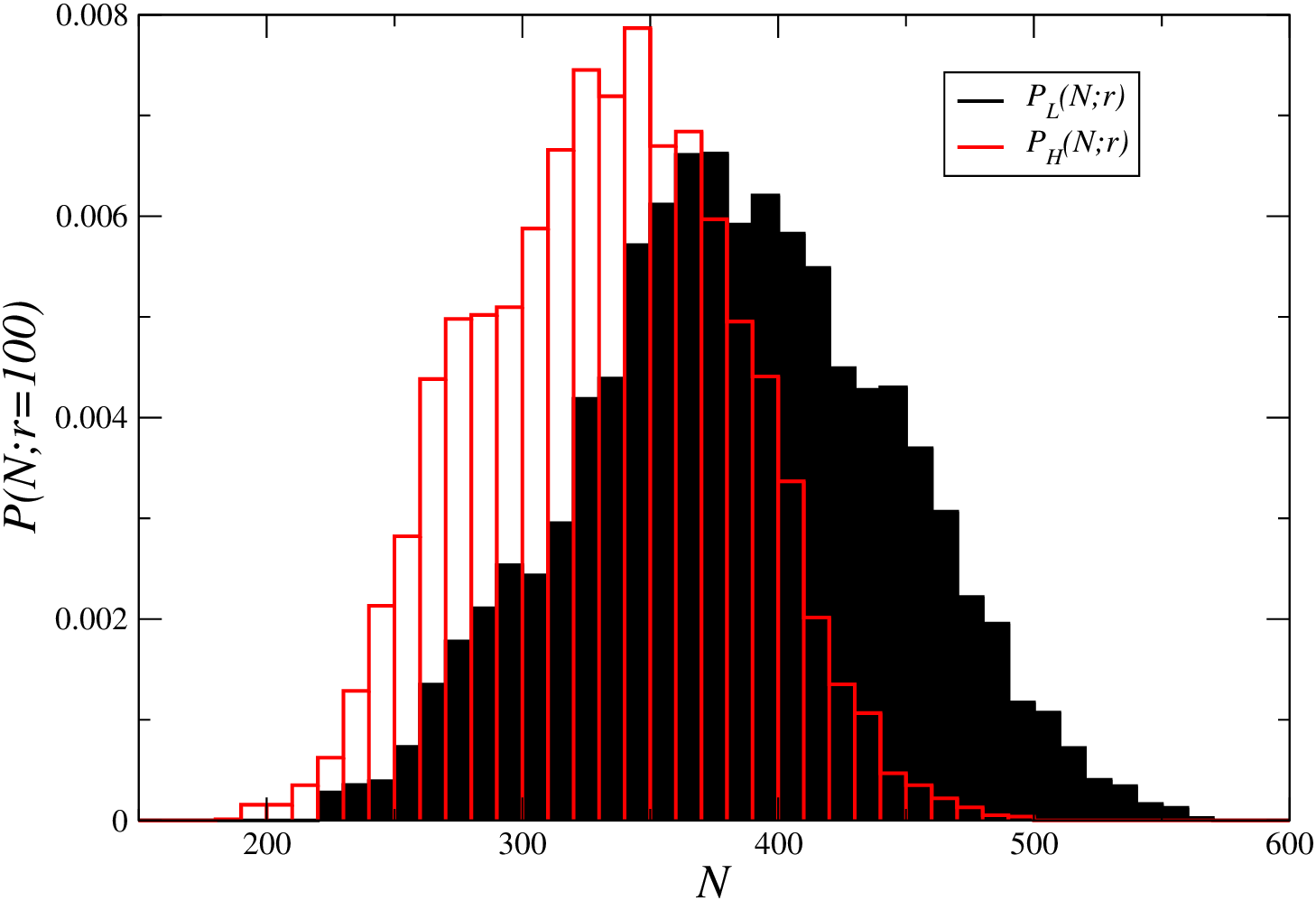}
\end{center}
\caption{As Fig.\ref{fig:pn2r50} but for $r=100$ Mpc/h. Here
  $M(r)=7929, 7690$.}
\label{fig:pn2r100}
\end{figure}

\begin{figure}
\begin{center}
\includegraphics*[angle=0, width=0.5\textwidth]{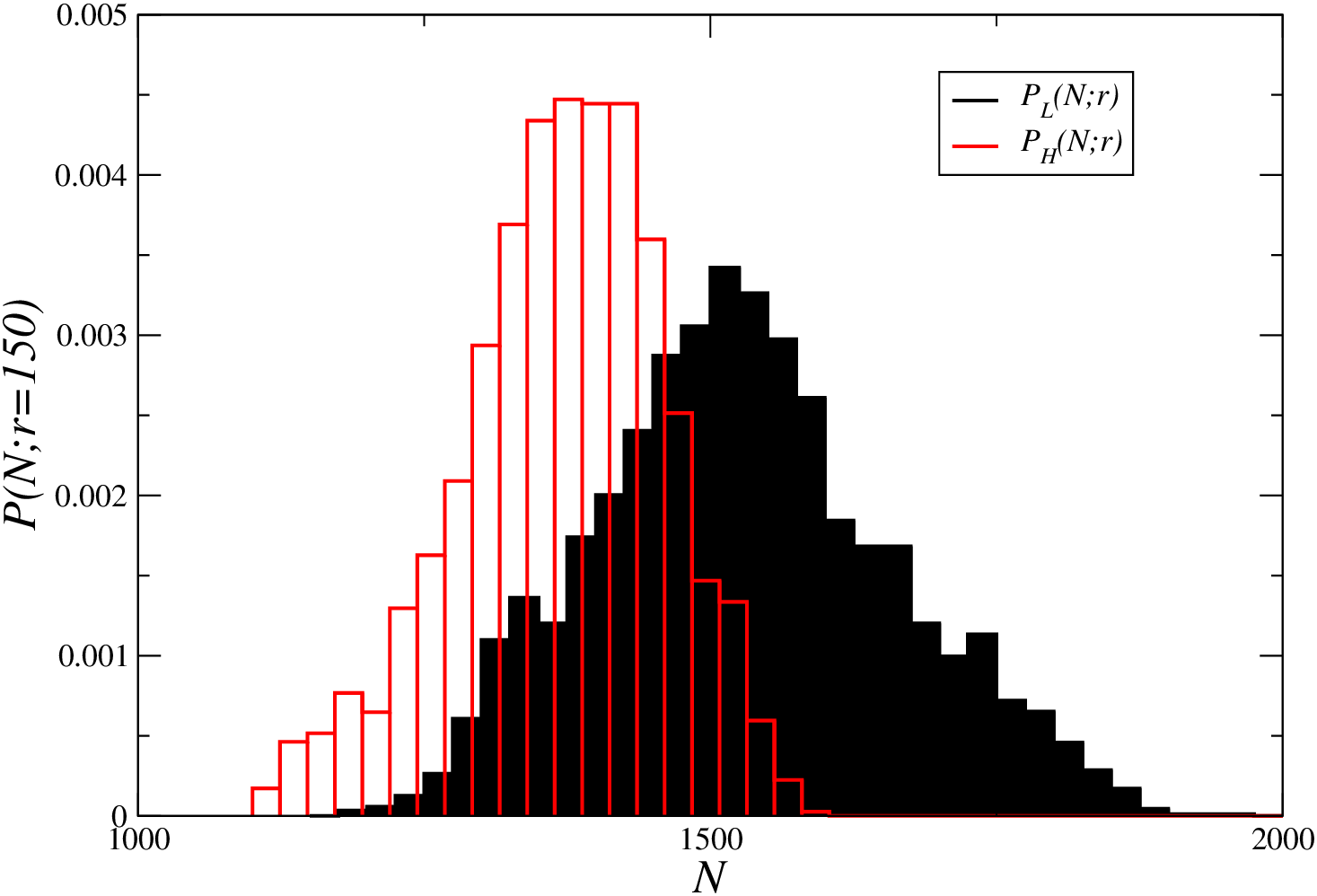}
\end{center}
\caption{ As Fig.\ref{fig:pn2r50} but for $r=150$ Mpc/h. Here
  $M(r)=3495, 3150$.}
\label{fig:pn2r150}
\end{figure}

\begin{figure}
\begin{center}
\includegraphics*[angle=0, width=0.5\textwidth]{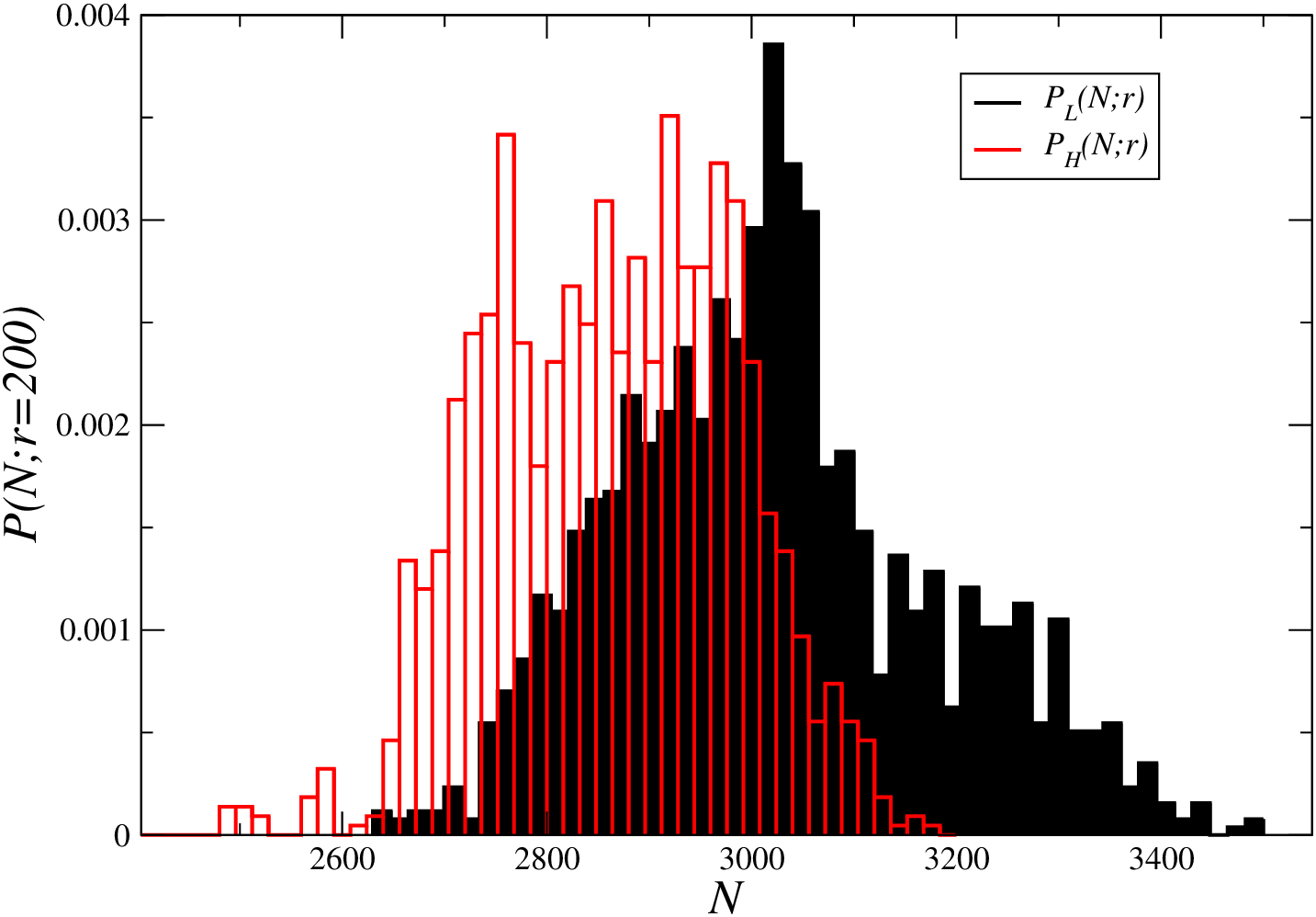}
\end{center}
\caption{As Fig.\ref{fig:pn2r50} but for $r=200$ Mpc/h. Here
  $M(r)=1465, 1354$.}
\label{fig:pn2r200}
\end{figure}

To clarify the features expected in LCDM models, we have done the same
analysis in realization of the LRG sample from the Horizon mock
samples \cite{horizon}. In
Figs.\ref{fig:pn2r100_mock}-\ref{fig:pn2r150_mock} we report the PDF
computed in the full sample $P(N;r)$ and in the two half-samples
$P_L(N;r)$ and $P_H(N;r)$ as done for the real sample.  One may note
that the PDF is much more regular and has a bell-shape.  Indeed the
Gaussian fit is reasonably good.  A complementary aspects of these
features is provided, as discussed above, by the behavior of the
radial density that is almost constant but noisy, where in this case
fluctuations are only due to Poisson noise and radial clustering (see
Fig.\ref{fig:radialden}).  These results are clearly expected from a
simple theoretical analysis: LCDM models are such that non-linear
clustering occurs only at small scales $r<10$ Mpc/h and thus the
distribution properties at $\sim 100$ Mpc/h are characterized by
Gaussianity, small amplitude fluctuations and weak two-point
correlations (see discussion in \citet{sdss_aea}).
\begin{figure}
\begin{center}
\includegraphics*[angle=0, width=0.5\textwidth]{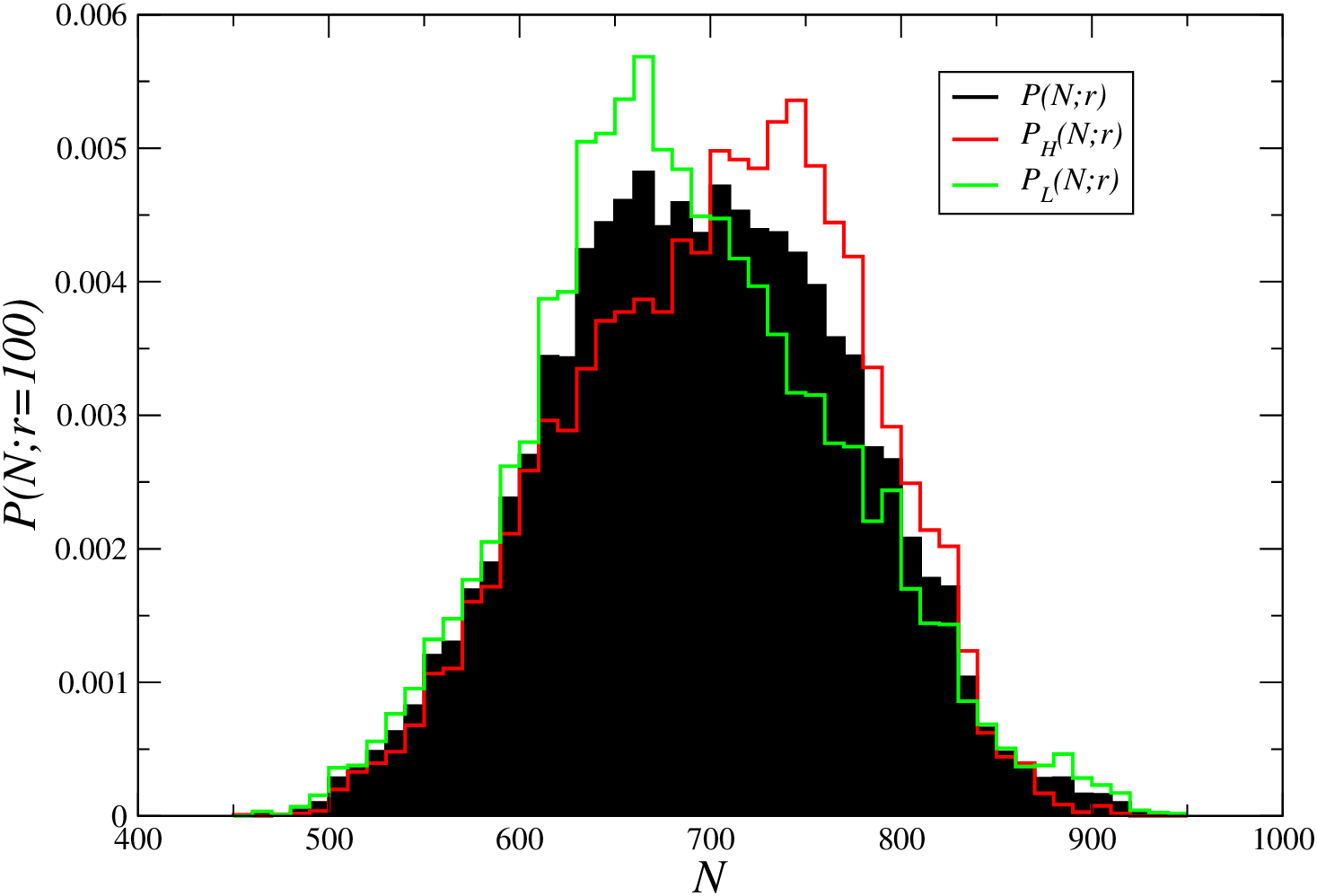}
\end{center}
\caption{PDF for $r=100$ Mpc/h in the low (L) redshifts and the high
  (H) redshifts sample for the Horizon mock LRG sample.}
\label{fig:pn2r100_mock}
\end{figure}
\begin{figure}
\begin{center}
\includegraphics*[angle=0, width=0.5\textwidth]{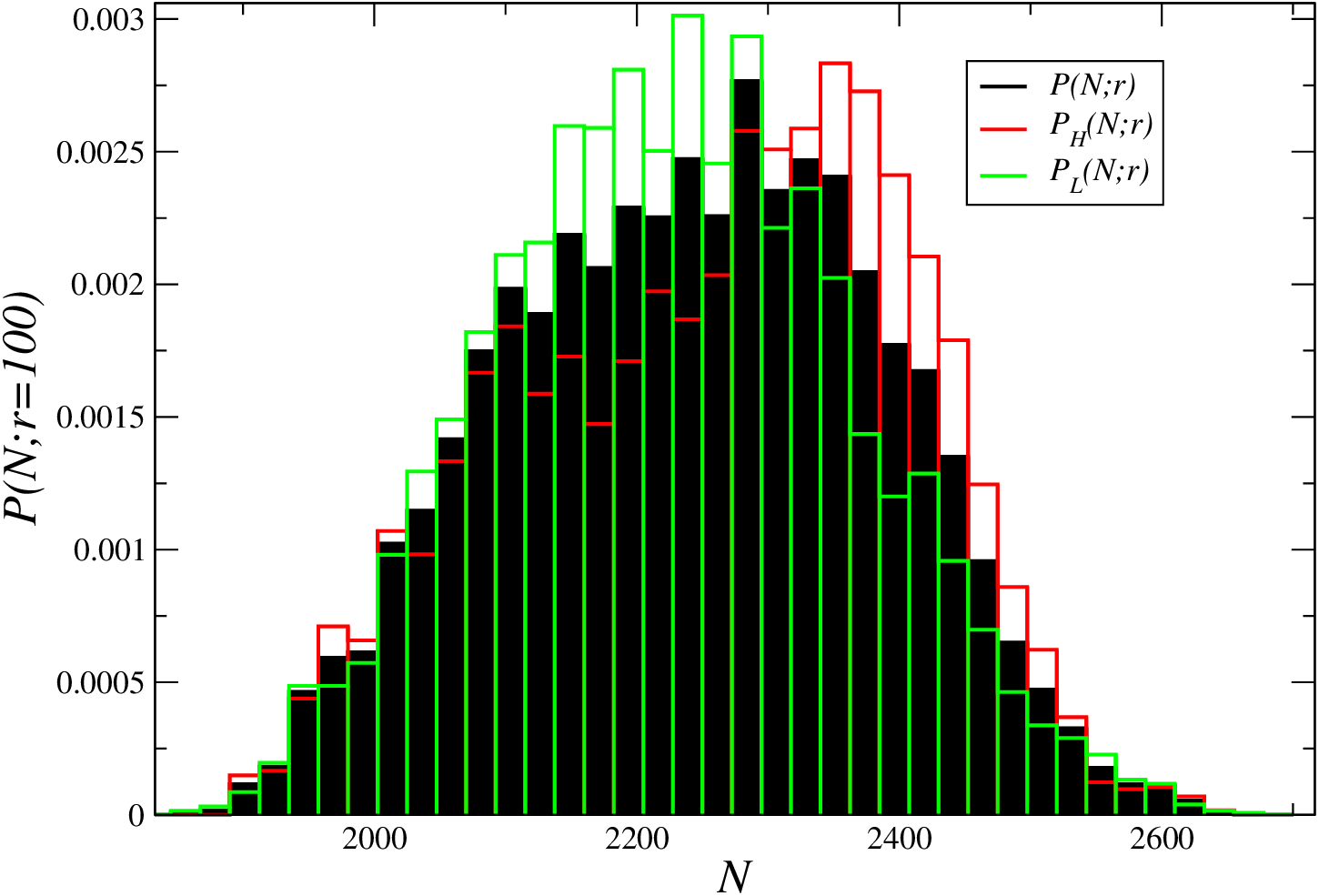}
\end{center}
\caption{PDF for $r=150$ Mpc/h in the low (L) redshifts and the high
  (H) redshifts sample for the Horizon mock LRG sample.}
\label{fig:pn2r150_mock}
\end{figure}

As a final remark, we note that when applying a random selection to
the mock catalog, in such a way to get the same $n(z)$ of the real
data, the results of our tests are compatible with the data. Thus, a
possible, but as mentioned, unsatisfactory, conclusion is that the
LRG is {\it compatible} with the mock catalog. We stress again that
this conclusion is based on the assumption that an important selection
effect, which is not smooth with redshift, occurs in the definition of
the sample.

\section{Discussion}
\label{discussion}

We have discussed that the number density as a function redshift
$n(z)$ of the LRG sample shows, for a wide redshift range, a
decreasing behavior as a function of redshift, followed by a sharp
increase around $z \sim 0.36$.  A different behavior was detected by
the $n(z)$ of the bright galaxies in the MGS. In particular
\citet{loveday} found that the number density of bright galaxies
increases by a factor $\approx 3$ as redshift increases from $z = 0$
to $z = 0.3$.  This is shown, for instance, by the increase of $n(R)$
up tp $z\sim 0.2$ in the two volume limited samples VL4 and VL5 of the
MGS shown in Fig.\ref{LRG+MGS_nr} (from the \citet{sdss_aea}). At
smaller scales the differential number density shows a fluctuating
behavior.  To explain the rise of the differential number density in
the deepest samples, which contain in average the brightest galaxies
of the MGS, a significant evolution in the luminosity and/or number
density of galaxies at redshifts $z < 0.3$ as been invoked
\citep{loveday}.  On the other hand, \citet{sdss_aea} by measuring the
PDF of galaxy counts in spheres, concluded that the volume-limited
samples of the MGS, extracted from the SDSS sixth data release, are
characterized by the breaking of self-averaging properties for scales
$r>30 $ Mpc/h and that the increase in the bright galaxies number
density is interpreted as an effect due to large scale
inhomogeneities.

\begin{figure}
\begin{center}
\includegraphics*[angle=0, width=0.5\textwidth]{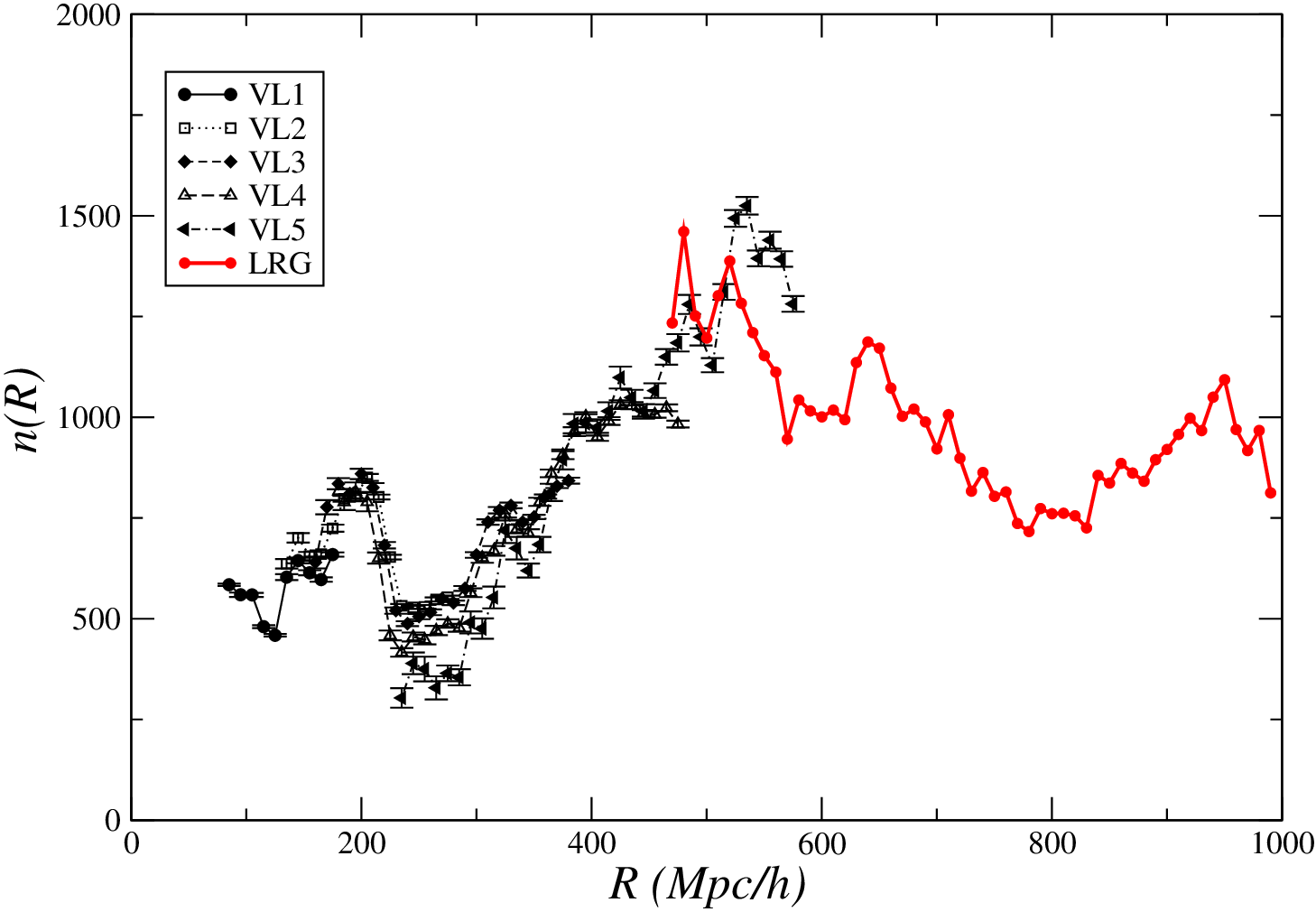}
\end{center}
\caption{Radial density in the volume limited samples of the MGS (from
  \citet{sdss_aea}) and in the LRG sample.  the $n(z)$ for the MGS
  volume limited samples has been normalized by taking into account
  the different selection in luminosity in the different samples while
  for the LRG units are arbitrary.}
\label{LRG+MGS_nr}
\end{figure}

In addition, for galaxies with magnitudes around $M^*$ it was
subsequently shown by \citet{tibor} that self-averaging properties are
verified when the larger volumes of the final data release of SDSS is
considered. This actually poofs that the breaking of self-averaging
properties is due to a finite volume effect and not to galaxy
evolution \citep{copernican}. Moreover in the deeper samples,
containing the brightest galaxies, self-averaging properties were
found not to be satisfied even in the SDSS final data release
\citep{copernican}. In that case, in agreement with \citet{loveday},
it was observed a growth of radial density rather than a decrease.  In
summary, these behaviors support the conclusion that galaxies from the
MGS of the SDSS are not compatible with spatial homogeneity at scales
$\sim 100$ Mpc/h. A similar conclusion was drawn from the analysis of
the 2dFGRS by \citet{2df_epl,2df_aea}. Recently \citet{saslaw2010}
found that cosmic variance in the SDSS causes the counts-in-cells
distributions in different quadrants to differ from each other by up
to 20\%, a result which corroborate the results discussed here.

We note that the same systematic effects discussed for the behavior of
conditional fluctuations similarly affect the determinations of the
two-point correlation function and clearly the detection of the
so-called baryon acoustic oscillations
\citep{eisenstein,kazin,martinez}.  It was indeed shown by
\citet{sdss_bao} that major systematic effects are present in the
estimators of $\xi(r)$ and that the variance estimated from the sample
itself is much larger than the one deduced from the analysis of mock
catalogs, making the detection of the baryon acoustic oscillations
statistically unreliable.

Instead of investigating the origin of the fluctuating behavior of
$n(z)$, \citet{kazin} focused their attention on the effect of the
radial counts on the determination of the two-point correlation
function $\xi(r)$.   In particular, they proposed mainly two different
tests to study what is the effect of $n(z)$ on the determination of
$\xi(r)$.  The first test consists in taking a mock LRG sample,
constructed from a cosmological N-body simulation of the LCDM model,
and by applying a redshift selection which randomly excludes points in
such a way that the resulting distribution has the same $n(z)$ of the
real sample. Then one can compare $\xi(r)$ obtained in the original
mock and in redshift-sampled mock. \citet{kazin} find that there is a
good agreement between the two.  This shows that the particular kind
of redshift-dependent random sampling considered for the given
distribution, does not alter the determination of the correlation
function. In other words this shows that, under the assumption that
the observed LRG sample is a realization of the mock LCDM simulation,
the $n(z)$ does not affect the result. However, if we want to test
whether the LRG sample has the same statistical properties of the mock
catalog, we cannot clearly proof (or disproof) this hypothesis by
assuming a priori that this is true.

 The second test consists in computing $\xi(r)$ in the data, by using
 a pair-counting method which measures two-point correlations by using
 a random distribution as reference and by comparing the number of
 data-data pairs with the number of data-random and/or random-random
 pairs \citep{dp83,lz}. Then, to include the variations of $n(z)$ in
 the data one considers a random sample that is not a purely Poisson
 distribution. Instead, the random sample is generated so that it has
 the same $n(z)$ behavior as in the data.  This means that in each
 redshift shell at distance $z$ and of thickness $\Delta z$ the
 distribution is Poisson with the same density $n(z)$ as observed.
 This test shows only that a random sample modified in such a way,
 does not alter the determination of $\xi(r)$ (as it is found by
 \citet{kazin}). Instead, it does not show that the features in the
 observed $n(z)$ behavior are taken into account by the fact that the
 random sample has the same features in the redshift counts.

The behavior of the average conditional properties is clearly
ill-defined as long as self-averaging properties are not found to be
stable: indeed the average conditional density is just the first
moment of the PDF. In \citet{sdss_aea} we have discussed that up to
$\sim 30$ Mpc/h the average conditional density shows approximately a
$\sim r^{-1}$ behavior, while in \citet{tibor} we find that there is
an apparent change of slope at larger scales where the average
conditional density decays as $\sim r^{-0.3}$ approximately up to
$100$ Mpc/h. A change of slope at $\sim 30$ Mpc/h from $\sim r^{-1}$
was found by \citet{hogg} in a preliminary sample of the LRG. However
they claimed that a clear transition to a flat behavior, signaling
uniformity, was reached at $\sim 70$ Mpc/h. With the present data,
which represent an improvement of almost a factor 3 in volume with
respect to the data considered by \citet{hogg}, because of the
breaking of self-averaging properties, we are not able to make a
reliable determination of the average conditional density at scales of
the order of $\sim 100$ Mpc/h. However, we note that our results are
not compatible with a transition to uniformity at those scales.  

\section{Conclusions}
\label{conclusions}

The number density as a function of redshift in the final LRG sample
from the SDSS \citep{kazin} shows an irregular and mainly decaying
behavior which cannot be simply explained by a smooth
redshift-dependent selection. We have discussed that the behavior of
$n(z)$ is closely related to the properties of fluctuations in this
sample. We conclude that these are not self-averaging and thus they do
not allow us to make a precise statement on the behavior of volume
average quantities, as the average conditional density or the
two-point correlation function.  Either the behavior of $n(z)$ is
determined by intrinsic fluctuations in the LRG distribution which are
in amplitude and spatial extension much larger than the ones expected
in LCDM models, or there are major observational selection effects
intervening in the definition of the LRG sample.  For this reason we
conclude that forthcoming galaxy samples, as the data from the Baryon
Oscillation Spectroscopic Survey \citep{boss}, with more controlled
selection effects, will hopefully clarify the situation.

\acknowledgements I am grateful to Yuri V. Baryshev, Andrea Gabrielli,
Michael Joyce, Martin L\'opez-Corredoira and Nickolay L. Vasilyev for
useful collaborations and suggestions.  I also wish to thank David
Hogg, Abhilash Mishra and Subir Sarkar for discussions and comments.
I acknowledge the use of the Sloan Digital Sky Survey data ({\tt
  http://www.sdss.org}).



{}


\end{document}